\documentclass[12pt]{article}
\usepackage[english]{babel}
\usepackage{amsfonts}
\usepackage{amsmath}
\usepackage{amssymb}
\usepackage{mathrsfs}
\usepackage{latexsym}
\usepackage{multicol}
\usepackage{fancybox}
\usepackage{color}
\usepackage{hyperref}
\usepackage{graphicx} 
\usepackage{caption}

\def\be{\begin{equation}}
\def\ee{\end{equation}}
\def\tl{\tilde} 
 
\def\gm{\gamma} 
\def\lm{\lambda}

\def\d'{``}

\def\ba{{\bf a}} 
\def\bx{{\bf x}} 
\def\C{\mathbb{C}} 
\def\R{\mathbb{R}}
\def\Q{\mathbb{Q}}

\newtheorem{thm}{Theorem}[section]

\newtheorem{rem}[thm]{Remark}

\newtheorem{defn}[thm]{Definition}

\DeclareMathOperator*{\Max}{max}

\def\bea{\begin{eqnarray}}
\def\eea{\end{eqnarray}}

\def\i'{\textrm{i}}

\def\dblone{\hbox{$1\hskip -1.2pt\vrule depth 0pt height 1.6ex width 0.7pt \vrule depth 0pt height 0.3pt width 0.12em$}}
\def\d'{``}

\begin{document}

\begin{center}
\Large{\bf{Algebraic entropy for algebraic maps.}}
\end{center}

\begin{center}
{ {\bf A.N.W. Hone$^\dag$, Orlando Ragnisco$^\ddag$,  Federico Zullo$^\ddag$}}

{$^\dag$SMSAS, University of Kent, Canterbury, U.K.\\$^\ddag$Dipartimento di Matematica e Fisica, Universit\`a di Roma Tre \& \\ Istituto Nazionale di
Fisica Nucleare, sezione di Roma Tre, Roma, Italy\\
}
\end{center}

\medskip
\medskip





\medskip
\medskip

\begin{abstract}
\noindent
We propose an extension of the concept of algebraic entropy, as introduced by Bellon and Viallet for rational maps, 
to algebraic maps (or correspondences) of a certain kind. The corresponding entropy is an index of the complexity of the map. The definition inherits the basic properties 
from the definition of entropy for rational maps. We give an example with positive entropy, as well as two examples taken from the theory of B\"acklund transformations.
\end{abstract}

\bigskip\bigskip

\noindent


\section{Introduction} \label{intro}
The notion of algebraic entropy was introduced to detect the integrability of discrete dynamical systems represented by rational transformations. The idea was simple: looking at the growth of the degree of the iterates of the map, exponential growth is a strong evidence for chaos, while polynomial growth means regular behaviour. Usually a transformation $\phi$ of $m$-dimensional projective space $\mathbb{P}^m$ is written in terms of $m+1$ homogeneous coordinates; then the map is given by a set of $m+1$ homogeneous polynomials of the same degree $d$. If there are no factorizations, the degrees $d_n$ of the iterates $\phi^n$ are given by $d_n=d^n$. However, if factorizations occur then the values of the degrees will drop down: this reduced growth in the complexity of the iterates is in general a good predictor of the integrability of the map \cite{BV,HV}; indeed, the factorization process has been used as a classification tool for  integrable rational maps \cite{HV}. 
The algebraic entropy associated to rational maps is defined to be \cite{BV}
$$
\mathcal{E}=\lim_{n\to\infty}\frac{\log(d_n)}{n}. 
$$ 
The entropy can also be calculated for a rational map written in affine coordinates, in which case the degree of the map 
is the maximum of the degrees of the $m$ rational functions which define it. 
An entropy equal to zero means a  sub-exponential (usually polynomial) growth of the degrees. 
Many of the discrete integrable maps considered in the literature are not only rational but also birational, meaning that they have an inverse which is itself a rational map. 
Up to now a large number of birational maps have been analyzed; the two-dimensional case is the most thoroughly investigated (see e.g. \cite{V} and references therein), but there are also many results in three dimensions \cite{L1}.  

In this paper we propose a means to apply the algebraic entropy machinery to more complex situations, that is to algebraic maps or correspondences. 
Up until now, due to multivaluedness, the notion of algebraic entropy has not been defined for algebraic maps, although Veselov has suggested 
that a sub-exponential growth in the number of images is an indicator of integrability \cite{veselov}. 
The most common examples of algebraic maps 
come from the theory of B\"acklund transformations for integrable finite-dimensional Hamiltonian systems, which provided 
the original motivation for this work. Indeed, as will be seen in the following, in that setting the map is naturally defined on the field extension 
$\mathbb{C}(\bx)[\beta]$, where $\beta$ is a primitive element, satisfying 
a polynomial equation of some degree $N$: it depends on the dynamical variables $\bx = (x_1,x_2,\ldots,x_m)$ in an $m$-dimensional phase space, 
and may also depend on additional parameters of the transformation; in the case of B\"acklund transformations for systems defined by a Lax representation, it is related to the characteristic equation  of the Lax matrix (the spectral curve). It turns out that in this setting the primitive element is an algebraic function of the conserved quantities, so that the map is birational on any fixed orbit of the system. However, we shall see that our 
definition of algebraic entropy is applicable to a wider class of maps than just  B\"acklund transformations. 

The paper is organized as follows. In section \ref{sec1} we describe how to associate an algebraic entropy to a certain class of 
maps whose components are defined in a field extension $\mathbb{C}(\bx)[\beta]$. 
In general, algebraic maps are multivalued (they are correspondences), and the components of their higher iterates are not defined within the 
original field extension, but to apply our approach we restrict ourselves to 
the class of maps where all iterates can be expressed in $\mathbb{C}(\bx)[\beta]$.
In section 3 we consider two examples of non-integrable algebraic maps: the first fits into 
our scheme, and allows us to calculate the algebraic entropy, but the second does not.  
B\"acklund transformations (BTs) 
provide the most common examples of algebraic maps of the appropriate kind, and in section 4 we describe the general setting 
for these, while  
in section \ref{sec2} we shall give two examples of BTs, starting from a simple linearizable set of transformations 
(a discrete harmonic oscillator) and then considering the BT for a Hamiltonian with a cubic potential. 
The final section is reserved for some conclusions. 

\section{Algebraic entropy for algebraic maps}\label{sec1}
Let us consider the field $\C(\bx)$ 
of rational functions of the dynamical variables $\bx$ in an $m$-dimensional phase space. 
For example, when $m$ is even,  $\bx$ could be the set of canonically conjugate variables of a Hamiltonian mechanical system. The ground field 
is taken to be $\C$, the field of complex numbers, but in particular examples we could restrict this to be $\R$ or $\Q$, or extend it 
to the field of rational functions $\C ({\bf \mu})$,  
where  ${\bf \mu}$ indicates a set of parameters on which the transformations considered depend. Next, we consider $\C(\bx)[\beta]$,  
a simple extension of $\C(\bx)$ of degree $N$, 
where $\beta$ satisfies an irreducible polynomial equation of degree $N$ defined over $\C(\bx)$, that is 
\be\label{PN} 
P_N(\beta)\equiv \beta^N +\sum_{k=0}^{N-1} c_k(\bx) \beta^k=0, 
\ee 
where the coefficients $c_k$ are rational functions of $\bx = (x_1,x_2,\ldots,x_m)$. 
Now $\C(\bx)[\beta]$ is a vector space of dimension $N$ over $\C(\bx)$, meaning that 
any $f\in\C(\bx)[\beta]$ can be written as 
\be\label{obasis}
f = \sum_{j=1}^{N} \hat{a}_j \beta^{(j)},
\ee
where $\hat{a}_j\in  \C(\bx)$ and the independent elements $\beta^{(1)},\beta^{(2)},\ldots, \beta^{(N)}$ form a basis \cite{weyl}. 
For the purposes of our 
discussion it will be convenient to take the fixed basis $1,\beta,\ldots,\beta^{N-1}$ and write  
$$
\C(\bx)[\beta]=\left\{a_0+a_1\beta +\cdots +a_{N-1}\beta^{N-1}\,\left|\, a_k\in\C(\bx)\right.\right\}.
$$
However, the definition of entropy presented below is independent of the choice of basis. 

We now consider an $m$-dimensional map $\phi : \, \C^m\to \C^m$ whose components are defined over $\C(\bx)[\beta]$, given by 
\be\label{eqs} 
\phi: \quad \bx \mapsto \tl{\bx} = \sum_{j=0}^{N-1} \ba_j (\bx )\beta^j, 
 \ee 
where each coefficient $\ba_j$ is an $m$-component vector of rational functions in $\C(\bx)$, and 
$\beta = \beta(\bx)$ satisfies (\ref{PN}). Note that, in general, $\phi$ defined by (\ref{eqs}) is multivalued, so 
is not really a ``map'' as such; it is a correspondence of degree $N$, since from (\ref{PN}) 
there are $N$ different choices of $\beta$. If we wish to measure the growth of $\phi$ under iteration, then 
we immediately encounter the problem that the number of images proliferates (typically there are $N^n$ images after $n$ steps), 
and what is worse, generically the components of 
$\phi^2(\bx )$ and all subsequent iterates are not defined over $\C(\bx)[\beta]$, but require a further field extension at  
each step. In order to avoid the latter problem, we restrict ourselves to maps which satisfy the condition 
that $\tl{\beta}=\beta (\tl{\bx})=\beta (\phi(\bx))\in\C(\bx)[\beta]$, so that we may write 
\be\label{eqs2} 
\tl{\beta} = \sum_{j=0}^{N-1} b_j (\bx )\beta^j, 
\ee 
for some coefficients $b_j\in\C(\bx)$. 
By definition, $\tl\beta$ satisfies the equation 
\be\label{tb} 
\tl{\beta}^N +\sum_{k=0}^{N-1} c_k(\tl{\bx}) \tl{\beta}^k=0, 
\ee 
with $\tl{\bx}=\phi (\bx)$ defined by (\ref{eqs}). 

Given an algebraic map (\ref{eqs}) satisfying the condition (\ref{eqs2}), we can write down a pair of recurrence relations 
that determine the iterates $\bx_n =\phi^n(\bx)$, with the initial condition $\bx_0=\bx$, namely     
\begin{equation}\label{neqs}\begin{split}
&\bx_{n}=\sum_{k=0}^{N-1}\ba_{k}(\bx_{n-1})\beta_{n-1}^k,\\
&\beta_{n}=\sum_{k=0}^{N-1}b_{k}(\bx_{n-1})\beta_{n-1}^k ,
\end{split}\end{equation} 
where $\beta_n =\beta (\bx_n)$. 
Since $\beta$ is defined by (\ref{PN}), we have
\begin{equation}\label{gm}
\beta_n^N +\sum_{k=0}^{N-1} c_k(\bx_n) \beta_n^k=0.
\end{equation} 
Due to the recurrence for $\beta_n$ in (\ref{neqs}), the powers of $\beta_n$ and the coefficents $\ba_{k}(\bx_{n})$ 
on the right-hand side of the equation for 
$\bx_{n}$ can be rewritten as elements of   $\C(\bx)[\beta]$, with $\beta_0=\beta$ and $\bx_0=\bx$, leading to expressions of the form 
\begin{equation}\label{neqs2}\begin{split}
&\bx_{n}=\sum_{k=0}^{N-1}\ba_{k,n}(\bx)\beta^k,\\
&\beta_{n}=\sum_{k=0}^{N-1}b_{k,n}(\bx)\beta^k ,
\end{split}\end{equation}
where each component of the vector $\ba_{k,n}$, and each coefficient $b_{k,n}$, belongs to $\C(\bx)$. Notice that the multivaluedness of $\beta_n$ 
in (\ref{neqs2}) only enters through the initial choice of $\beta$, one of the $N$ roots of (\ref{PN}). Now each component $a_{k,n}^{(j)}$ of $\ba_{k,n}$ is a 
rational function of $\bx$, with degree given by the maximum of the degrees of the numerator and denominator, and we define 
$$ 
\deg \ba_{k,n} := \Max\limits_{j\in\{1,\dots,m\}}(\deg a_{k,n}^{(j)}). 
$$
Then we define the algebraic entropy of the map (\ref{eqs}) as follows:
\begin{defn}\label{def}
For the map $\phi$,  the degree of the $n$th iterate $\bx_n$, as in (\ref{neqs2}), is defined to be
\be\label{degdef} 
d_n:=\Max\limits_{k\in\{0,\dots,N-1\}} (\deg \ba_{k,n}) 
\ee
Then the algebraic entropy of $\phi$ is 
\begin{equation}\label{entdef}
\mathcal{E}=\lim_{n\to\infty}\frac{\log(d_n)}{n}.
\end{equation}
\end{defn}
For $N=1$, the formula (\ref{degdef}) reduces to that for the degree of a rational map given in affine coordinates, and so the above definition coincides with that of the algebraic entropy for rational maps \cite{BV}.  
The definition of degree in (\ref{degdef}) is not preserved under change of basis in $\C(\bx)[\beta]$, considered as a vector space over $\C(\bx)$. However, if the elements of $\C(\bx)[\beta]$ are expanded with respect to a different basis, as in (\ref{obasis}), then in the limit the value 
of the entropy will remain the same. One could use various alternative definitions of degree, which are manifestly invariant under change of basis, 
by exploiting the algebra structure of  $\C(\bx)[\beta]$ (see the first chapter of \cite{weyl} for details). To be precise, in any given basis, 
multiplication by an element $f\in\C(\bx)[\beta]$ can be represented by a matrix $F=(F_{jk})$ whose elements belong to  $\C(\bx)$, i.e. for any 
$g=\sum \hat{b}_j\beta^{(j)}\in \C(\bx)[\beta]$ the product with $f$ can be expanded as 
$$ 
f\cdot g = \sum_{j,k=1}^N F_{jk}(\bx ) \hat{b}_k(\bx ) \beta^{(j)}.
$$ 
Then coefficients of the characteristic polynomial of $F$, and in particular the trace 
$$ 
\mathrm{S}(f) = \mathrm{tr}\, F, 
$$ 
and the norm 
$$ 
\mathrm{Nm}(f) = \det F, 
$$ 
are independent of the basis.   
\begin{rem}\label{rem1} As an alternative to the definition (\ref{degdef}),  the growth of $\phi$ could be measured by replacing $d_n$ with one of the 
quantities $d_{\mathrm{S},n}$ or $d_{\mathrm{Nm},n}$, given by the maximum of the traces/norms of the components of $\bx_n$, respectively.
Yet we expect that $d_{\mathrm{S},n} = d_n + O(1)$, and $ d_{\mathrm{Nm},n}= N d_n+O(1)$; so either of these alternatives should  
leave the value of $\mathcal{E}$ unchanged.  
\end{rem} 
There is another way to interpret the map  (\ref{eqs}) together with the condition (\ref{eqs2}): one can view $\beta$ as an additional 
coordinate adjoined to $\bx$, so that (\ref{eqs}) and (\ref{eqs2}) define a rational 
map in $(m+1)$-dimensional affine space. In that setting, the equation (\ref{PN}), defines a submanifold of 
codimension one, 
that is 
$$ 
\mathcal{M}_P:=\{ (\bx,\beta)\in\C^{m+1} \, \vert \, P_N(\beta)=0\}, 
$$ 
and it is an invariant submanifold whenever (\ref{tb}) holds for all $ (\bx,\beta)\in\mathcal{M}_P$. 
\begin{rem} \label{rem1b} 
Any rational map in $\C^{m+1}$ with a rational first integral $H$ defines an algebraic map in $\C^m$, of the required 
form for Definition \ref{def} to be applied. Indeed, if 
we isolate one of the coordinates, denoted $\beta$, then the map for $(\bx, \beta)\in\C^{m+1}$ can always be written 
in the form of (\ref{eqs}) coupled with a condition of the form (\ref{eqs2}), by restricting to a fixed level set 
$H(\bx,\beta)=h$ (and clearing denominators where necessary). 
Each level set is an invariant submanifold of codimension one, defined by the vanishing of a polynomial in $\beta$ of the form (\ref{PN}), 
for some $N$. 
\end{rem}
In the next section, by exploiting the preceding remark, we present an example of an algebraic map for which Definition \ref{def} 
is relevant, and give another example where our approach cannot be applied. 
Algebraic recurrences of the form  (\ref{neqs}) also  arise naturally via iteration of B\"acklund transformations, as will 
be explained subsequently. 

\section{Non-integrable examples in two dimensions}\label{nonint}  
We now consider two different examples of algebraic maps in dimension $m=2$. In both cases, we have a field extension of degree 
$N=2$. For the first example, we are able to obtain a condition of the form  (\ref{eqs2}), and use this to calculate the degrees 
of the iterates according to (\ref{degdef}), and hence find the algebraic entropy. However, for the second example we prove that 
(\ref{eqs2}) cannot hold, so  Definition \ref{def} does not apply. 

\subsection{An example from a trace map} 

Starting from the algebraic map 
\be\label{tracem} 
\left(\begin{array}{c} x\\ y \end{array}\right) \, \mapsto\,  
\left(\begin{array}{c} \tl{x}\\ \tl{y} \end{array}\right) = \left(\begin{array}{c} y\\ \beta \end{array}\right),   
\ee 
with 
\be\label{quad} 
\beta^2 -axy\beta +x^2+y^2+b=0 
\ee  
where $a,b$ are arbitrary parameters, 
we seek a condition of the form  (\ref{eqs2}). Since the equation (\ref{quad}) is of degree 2 in $\beta$, we 
should have 
\be\label{bup}
\tl{\beta} = b_0(x,y) +b_1(x,y) \beta, 
\ee  
where $b_0$ and $b_1$ are rational functions of their arguments. Then $\tl{\beta}$ should also satisfy 
$$
\tl{\beta}^2 -a\tl{x}\tl{y}\tl{\beta} +\tl{x}^2+\tl{y}^2+b= 0, 
$$ 
and, upon using (\ref{tracem}), this becomes
\be\label{quads} 
\tl{\beta}^2 -ay\beta\tl{\beta} +y^2+\beta^2+b= 0.
\ee 
Now, by subtracting (\ref{quad}) from (\ref{quads}), we find 
\be\label{roots} 
(\tl{\beta} -x)( \tl{\beta} -ay\beta +x)=0.
\ee 
There are two different choices for $\tl\beta$ as elements of $\C(x,y)[\beta]$, but taking the first factor in (\ref{roots}) 
leads to a map with all orbits having period 3, so we take the second choice, that is 
\be\label{bcond} 
\tl{\beta}=ay\beta -x, 
\ee 
and hence we have $b_0=-x$ and $b_1=ay$. 

Iteration of the map (\ref{tracem}), together with the condition (\ref{bcond}), generates the recurrence relations 
\be \label{trecs} 
\begin{array}{rcl} 
x_n & = & y_{n-1}, \\ 
y_n & = & \beta_{n-1}, \\ 
\beta_n & = &  ay_{n-1}\beta_{n-1} -x_{n-1} 
\end{array} 
\ee
for $n\geq 1$, with initial conditions 
$$ 
x_0=x, \qquad y_0=y, \qquad \beta_0 =\beta = \beta (x,y).
$$ 
Clearly $x_n$ belongs to $\C(x,y)[\beta]$,  so we have 
\be\label{xns} 
x_n = a_{0,n}(x,y)  + a_{1,n}(x,y) \beta 
\ee 
for suitable rational functions (in fact, polynomials) $a_{0,n}$ and $a_{1,n}$; and since $y_n=x_{n-1}$ it is sufficient to consider 
$x_n$ alone and calculate 
the degree $d_n =\deg x_n =\max (\deg a_{0,n}, \deg a_{1,n})$. Now $x_0=x$, $x_1=y$, $x_2=\beta$, giving 
$d_0=1=d_1$ and $d_2=0$, while empirically for $n\geq 3$ we find that $\deg a_{0,n}= \deg a_{1,n}=d_n$ and the 
sequence of degrees goes 
$$ 
1,3,6,11,19,32,53,87,142,231,375,\ldots , 
$$ 
so by inspection we see that the differences are Fibonacci numbers, i.e. 
$$ 
d_{n+1}-d_n = f_n\quad \mathrm{for}\quad n\geq 2, 
$$ 
with $f_1=f_2=1$, $f_{n+2}=f_{n+1}+f_n$. This gives a linear  recurrence of third order for the degrees, that is 
$$ 
d_{n+3}-2d_{n+2}+d_n=0, 
$$ 
with characteristic equation  $(\zeta -1)(\zeta^2-\zeta-1)=0$, 
so $d_n\sim C\zeta^n$ with $C>0$ and $\zeta$ being the golden mean, and the entropy is 
\be\label{loggm} 
\mathcal{E}=\log\left(\frac{1+\sqrt{5}}{2}\right). 
\ee

The algebraic map (\ref{tracem}) arises by applying the construction in Remark \ref{rem1b} to the birational map 
\be\label{fibm} 
\left(\begin{array}{c} x \\ y \\ z \end{array} \right) \mapsto 
\left(\begin{array}{c}  y \\ z \\ ayz - x\end{array} \right),  
\ee  
which is the Fibonacci trace map appearing in the theory of quasicrystals (see \cite{rb94} and references for details). This 
map is volume-preserving (but orientation-reversing) and has the first integral 
$$ 
H=axyz -x^2-y^2-z^2, 
$$ 
from which one sees that (\ref{tracem}) arises by restricting to a fixed level set $H=b$ and eliminating the variable 
$z=\beta$. 
For comparison, it is interesting to apply the standard algebraic entropy test directly to the map (\ref{fibm}), whose 
iterates are equivalent to those of the recurrence 
\be\label{fibtrec}
x_{n+3}=ax_{n+1}x_{n+2}-x_n. 
\ee 
All of the iterates are polynomials in the initial values $x_0=x$, $x_1=y$, $x_2=z$, and for $n\geq 0$ the sequence 
of degrees $\hat{d}_n =\deg x_n$ begins 
$$ 
1,1,1,2,3,5,8,13,21,34,55,89,144,\ldots.
$$ 
For $n\geq 1$  this is the Fibonacci sequence, as can be proved directly from (\ref{fibtrec}) by noting that 
$$ 
\hat{d}_{n+3} = \max(  \hat{d}_{n+1} +\hat{d}_{n+2}, \hat{d}_{n})
$$ 
and using the fact that $\hat{d}_{n+1} +\hat{d}_{n+2}> \hat{d}_{n}$, 
so the Fibonacci recurrence $\hat{d}_{n+3} = \hat{d}_{n+2} +\hat{d}_{n+1}$ holds  for all $n\geq 0$. 
This immediately implies that the algebraic entropy is the logarithm of the golden mean, in agreement with the value (\ref{loggm}) 
found for the corresponding algebraic map (\ref{tracem}). 

The map (\ref{fibm}) appears in various other contexts. For $a=3$ it can be used to generate sequences of Markoff numbers, 
which appear in Diophantine approximation theory. On a fixed level set of the first integral $H$ it is equivalent 
to iteration of the recurrence 
$$ 
x_{n+3}x_n = x_{n+2}^2 + x_{n+1}^2 + b, 
$$ 
which has the Laurent property (and arises from a cluster algebra when $b=0$). One of us has considered orbits defined over $\Q$  \cite{hone},  
applying the Diophantine integrability test proposed in \cite{halburd}, to find that generically the logarithmic heights $h_n$
of iterates grow at the same rate as the degrees, so that $(\log h_n )/n$ tends to the same limit  (\ref{loggm}).

\subsection{The Cohen map} 

The algebraic map 
\be\label{cohen} 
\left(\begin{array}{c} x\\ y \end{array}\right) \, \mapsto\,  
\left(\begin{array}{c} \tl{x}\\ \tl{y} \end{array}\right) = \left(\begin{array}{c} y\\ \beta - x \end{array}\right),   
\ee 
with 
\be\label{bco} 
\beta^2 -y^2-\epsilon^2=0, 
\ee  
is known as the Cohen map, with the positive choice of square root $\beta=\sqrt{y^2+\epsilon^2}$ taken for 
$(x,y)\in\R^2$. It is a symplectic map of standard type. 
The parameter $\epsilon\neq 0$ is inessential, as it can be removed by rescaling. This map is very interesting because 
numerical experiments suggest that it should be integrable, with the real orbits being closed curves that foliate the plane. 
However, it has been shown that this map has no 
algebraic first integral \cite{cohenmap}, and it is not one of the maps of standard type admitting an analytic first integral 
classified by Suris \cite{suris}. Nevertheless, there is an echo of integrability coming from the limit $\epsilon\to 0$, which 
gives the recurrence 
$$ 
x_{n+2} + x_n =|x_{n+1}|.
$$ 
For real  orbits, upon noting that $|x|=\max (x,-x)$, the latter is seen to be ``tropical'' (it is defined in the 
$(\max, +)$ algebra), and it turns out that for this tropical map all orbits are periodic with period 9 \cite{period9}. 

If we would wish to apply   Definition \ref{def} to (\ref{cohen}), then a condition of the form 
\be\label{btry} 
\tl{\beta} = p(x,y)+q(x,y)\beta 
\ee 
is required, for rational functions $p,q\in  \C(x,y)$. From (\ref{bco}), we must also have 
\be\label{bcos} 
\tl{\beta}^2 -(\beta-x)^2-\epsilon^2=0.
\ee 
Substituting for $\tl{\beta}$ from (\ref{btry}) and using (\ref{bco}) to eliminate terms in $\beta^2$  
from (\ref{bcos}), we obtain the equation 
\be\label{pq} 
(q^2-1)(y^2+\epsilon^2)+p^2-x^2 -\epsilon^2+ 2(pq+x)\beta = 0.
\ee
For this equation told in $\C(x,y)[\beta]$, the coefficients of both $\beta^0$ and $\beta^1$ must vanish. 
The coefficient of $\beta$ in (\ref{pq}) gives 
$$ 
q=-\frac{x}{p},
$$ and 
substituting this back into the other coefficient gives the quartic equation 
$$ 
p^4-( x^2 +y^2+2\epsilon^2)p^2 +x^2(y^2+\epsilon^2)=0
$$  
for $p$, which contradicts the assumption that $p\in\C(x,y)$. This shows that 
Definition \ref{def} cannot be applied to (\ref{cohen}), because the components of higher 
iterates of this map are not defined in  $\C(x,y)[\beta]$. 

\section{Algebraic maps from B\"acklund transformations}\label{bts} 
We consider a finite-dimensional integrable system described by a Lax matrix $L(\bx,\lm)$; 
the elements of the Lax matrix, 
$L_{i,j}(\lm)$ for $i,j \in \{ 1,\ldots ,N\}$, are functions of the dynamical variables $\bx$ and of the 
spectral parameter $\lm$. The transformed variables $\tl{\bx}$ given by a B\"acklund transformation define a new Lax matrix $L(\tl{\bx},\lm)$. 
Upon iteration, we can identify $\bx$ and ${\tl{\bx}}$ with ${\bx}_{n-1}$ and ${\bx}_n$ in (\ref{neqs}), respectively. 
In order to preserve the spectrum of the Lax matrix, which corresponds to the set of commuting first integrals of the system, 
the matrices $L({\tl{\bx}},\lm)$ and $L({\bx},\lm)$ must be similar, i.e. associated to any transformation there is a matrix $D(\lm)$ such that 
\begin{equation}\label{LtDDL}
L({\tl{\bx}},\lm)D(\lm)=D(\lm)L(\bx,\lm)
\end{equation}
(see e.g. \cite{KS,KV,RZ1,RZ,Z}). 
One way to obtain B\"acklund transformations is to construct such a matrix $D(\lm)$. 
Obviously $L(\tl{\bx},\lm)$ and $L(\bx,\lm)$ must have the same structure in $\lm$ 
and this constrains the dependence of $D(\lm)$ on the spectral parameter. Furthermore,  $D(\lm)$ will depend on the dynamical variables. 
This dependence can be of two types: either explicit or through a set of auxiliary variables. Suppose that the matrix $D(\lm)$ is singular when $\lm$ is equal to a particular value, say $\mu$. For the sake of simplicity, we also assume that the equation $\det(D(\mu))=0$ leaves just one auxiliary variable undetermined. We call this variable $\beta$. Since $D(\mu)$ is singular, it possesses a kernel, so for some non-zero $|\Omega\rangle$ we have 
$$
D(\mu)|\Omega\rangle=0.
$$
The elements of the vector $|\Omega\rangle$ depend on the dynamical variables, on the auxiliary variable $\beta$, and on $\mu$. We can choose a normalization by fixing one of the elements of the vector $|\Omega\rangle$ to be equal to 1, say the $k$th element. From equation 
(\ref{LtDDL}) it follows that $|\Omega\rangle$ is also an eigenvector of $L(\mu)$, so that 
$$
L(\bx,\mu)|\Omega\rangle=\gm(\mu) |\Omega\rangle, 
$$ 
where $\gm(\mu)$ is the corresponding eigenvalue. Note that, due to the normalization of $|\Omega\rangle$, 
the eigenvalue $\gm(\mu)$ is equal to $\sum_j L_{k,j}|\Omega\rangle_j$, so from the characteristic equation it follows that 
\begin{equation}\label{alg}
\det\left(L(\mu)-\sum_j L_{k,j}|\Omega\rangle_j\dblone\right)=0.
\end{equation} 
If the dependence of the elements of $|\Omega\rangle$ is polynomial in $\beta$ then the previous equation 
is a polynomial in $\beta$ and the degree is at least $N$. Indeed, to the best of our knowledge, 
the degree is exactly $N$ for all B\"acklund transformations of finite-dimensional integrable systems, 
since the elements of the kernel $|\Omega\rangle$ are linear in $\beta$ (see e.g. \cite{HKR1,HKR,KV,RZ1,RZ,Z1,Z,Z2}). 
Equation (\ref{alg}) then corresponds to equation  (\ref{gm}) for the variable $\beta$. 
The transformation for $\bx$ in (\ref{eqs}) is determined from (\ref{LtDDL}).
\begin{rem}
Since the spectrum of the Lax matrix 
is preserved by (\ref{LtDDL}), this means that $\gm(\mu)$ is independent of $n$, or, in other words $\gm(\mu,\tl{\bx})=\gm(\mu,\bx)$. 
The equation (\ref{alg}) is equivalent to
$$
\det\left(L(\mu)-\gm\dblone\right)=0,
$$
which is a polynomial in $\gm$ of degree $N$, defining a curve in the $(\mu,\gm)$ plane - the spectral curve. 
\end{rem}

\begin{rem}
In the case of B\"acklund transformations, because $\gm$ is a conserved quantity, 
it is more convenient to view the components of the map as being defined over $\C(\bx)[\gm]$ rather than $\C(\bx)[\beta]$.
Thus one can replace  $\beta_n$ with $\gm_n$ throughout the formulae for the iterates, so that the second equation of (\ref{neqs}) 
just becomes 
\begin{equation}\label{backform}
\gm_n=\gm_{n-1},
\end{equation}
and (\ref{gm}), which is the equation for the spectral curve, reads as  
\begin{equation}\label{backform1}
\gm^N +\sum_{k=0}^{N-1} c_k(\bx_n,\mu) \gm^k=0
\end{equation} 
for all $n$, where the coefficients $c_k(\bx_n,\mu)$ are conserved quantities (independent of $n$). 
The inverse map is automatically defined over $\C(\bx)[\gm]$, 
via the equation $\hat{D}(\lm)L(\tl{\bx},\lm)=L(\bx,\lm)\hat{D}(\lm)$ (see equation (\ref{LtDDL})), 
where $\hat{D}(\lm)$ is the adjugate of the matrix $D(\lm)$.
\end{rem}

In the next section we give some simple examples of algebraic maps from B\"acklund transformations, and calculate their algebraic entropies according to Definition (\ref{def}). Since the maps are integrable we find their entropies to be zero, as expected.

\section{Examples of B\"acklund transformations}\label{sec2}

We now calculate the algebraic entropy for two examples of algebraic maps coming from Hamiltonian systems with one degree of freedom. 
The first example, which is the B\"acklund transformation for the harmonic oscillator \cite{RZ}, is very simple but instructive since 
all the calculations can be made by hand. 
The second example comes from a Hamiltonian with a cubic potential, and is the simplest case of the discrete Mumford systems 
considered in \cite{KV}. 

\subsection{Discrete harmonic oscillator} 

The Lax representation with spectral parameter for the one-dimensional harmonic oscillator is given by 
\begin{equation*}
L(\lm) = \left(\begin{array}{cc} 1 & (p-\textrm{i}q)\lm^{-1}\\ (p+\textrm{i}q)\lm^{-1} & -1\end{array} 
\right) , \quad M = \frac{\textrm{i}}{2}\left(\begin{array}{cc} 1 & 0\\ 0& -1\end{array} \right), \quad \dot{L}(\lm)=[L,M]
\end{equation*}
The spectral curve, conserved under the action of the flow, is defined by
$$
\gm^2(\lm)=\det(L(\lm))=1+\frac{p^2+q^2}{\lm^2}.
$$
We shall write the transformations in the new variables 
$$
a:= p-\textrm{i} q, \qquad b= p+\textrm{i} q
$$  
A dressing matrix $D(\lm)$, providing a one-parameter family of B\"acklund transformations, is given by \cite{RZ}
\begin{equation*}
D(\lm) = \left(\begin{array}{cc} 1 & \mu\beta\lm^{-1} \\  \mu \beta^{-1}\lm^{-1}& 1\end{array} 
\right), \qquad  \beta:=-\frac{\mu (1+\gm(\mu))}{b} = \frac{a}{\mu(1-\gm(\mu))},
\end{equation*}
from which we find
\begin{equation*}\label{explBTs} \left.\begin{aligned}
&\tl{a}=a \frac{\gm(\mu) +1}{\gm(\mu) -1}\\
&\tl{b}=b \frac{\gm(\mu) -1}{\gm(\mu) +1}
\end{aligned}\right\} \quad \textrm{with}\quad \gm(\mu)^2 = 1+\frac{ab}{\mu^2}.
\end{equation*} 
The previous transformations define the recursions (cf. equations (\ref{backform}), (\ref{backform1}))
\begin{equation}\label{exp2BTs} 
a_{n}=a_{n-1} \frac{\gm_{n-1}+1}{\gm_{n-1}-1}, \qquad 
b_{n}=b_{n-1} \frac{\gm_{n-1}-1}{\gm_{n-1}+1}, 
\ee
and 
$$ 
\gm_{n}=\gm_{n-1},\qquad
\gm_n^2=1+\frac{a_nb_n}{\mu^2}
$$ 
for all $n$. The components of the map (\ref{explBTs}) and its iterates are  naturally defined over $\C(a,b)[\gm]$: 
each element $(a_k,b_k)$ can be expressed as a sum $(A_k,B_k)+\gm (C_k,D_k)$ (cf. equation (\ref{neqs})), where $A_k$, $B_k$, $C_k$ and $D_k$ are rational functions of $a$, $b$, and are also rational functions of the parameter $\mu$.

From formulae (\ref{exp2BTs}) we can find the functions $A_n$, $B_n$, $C_n$ and $D_n$ explicitly. Indeed, with $a_0=a$, $b_0=b$ and $\gm_0=\gm$, using the binomial theorem and the definition of $\gm$, we get
\begin{equation} 
\begin{aligned}
&a_{n}=a \left(\frac{\gm +1}{\gm -1}\right)^n=\frac{a\mu^{2n}}{a^nb^n}\sum_{k=0}^n{2n \choose 2k}\left(1+\frac{ab}{\mu^2}\right)^k+\gm \frac{a\mu^{2n}}{a^nb^n}\sum_{k=0}^{n-1}{2n \choose 2k+1}\left(1+\frac{ab}{\mu^2}\right)^k,\\
&b_{n}=b \left(\frac{\gm -1}{\gm +1}\right)^n=\frac{b\mu^{2n}}{a^nb^n}\sum_{k=0}^n{2n \choose 2k}\left(1+\frac{ab}{\mu^2}\right)^k-\gm \frac{b\mu^{2n}}{a^nb^n}\sum_{k=0}^{n-1}{2n \choose 2k+1}\left(1+\frac{ab}{\mu^2}\right)^k.
\end{aligned}
\end{equation} 
The rational functions $A_n$, $B_n$, $C_n$ and $D_n$ are then given by
\begin{equation}\begin{split}
&A_n=\frac{a\mu^{2n}}{a^nb^n}\sum_{k=0}^n{2n \choose 2k}\left(1+\frac{ab}{\mu^2}\right)^k, \quad B_n=\frac{b\mu^{2n}}{a^nb^n}\sum_{k=0}^n{2n \choose 2k}\left(1+\frac{ab}{\mu^2}\right)^k,\\
&C_n=\frac{a\mu^{2n}}{a^nb^n}\sum_{k=0}^{n-1}{2n \choose 2k+1}\left(1+\frac{ab}{\mu^2}\right)^k, \quad D_n=-\frac{b\mu^{2n}}{a^nb^n}\sum_{k=0}^{n-1}{2n \choose 2k+1}\left(1+\frac{ab}{\mu^2}\right)^k.
\end{split}\end{equation}
From these formulae it is clear that these rational functions are all of the same degree $2n$, so from Definition \ref{def} 
the algebraic entropy of the map (\ref{exp2BTs}) is  
$$
\mathcal{E}=\lim_{n\to\infty}\frac{\log (2n)}{n}=0.
$$

\subsection{Hamiltonian with a cubic potential} 

The Lax representation and the B\"acklund transformations for the Mumford systems are described in \cite{KV}. 
The simplest such system is defined by the Lax matrix
\begin{equation*}
L(\lm) = \left(\begin{array}{cc} p & 2(\lm^2+\lm q+q^2)\\ 2(\lm-q) & -p\end{array} 
\right),
\end{equation*}
and the $r$-matrix structure $ \{L(\lm),L(\eta)\}=[r_1,\stackrel{1}{L}+\stackrel{2}{L}]+[r_2,\stackrel{1}{L}-\stackrel{2}{L}]$, where
$$
r_1\doteq\frac{1}{\eta-\lm}\left( \begin{array}{cccc} 1&0&0&0\\ 0&0&1&0 \\ 0&1&0&0\\0&0&0& 1  \end{array} \right), \quad r_2\doteq\left( \begin{array}{cccc} 0&0&0&-1\\ 0&0&0&0 \\ 0&0&0&0\\0&0&0& 0  \end{array} \right),
$$
and $\stackrel{1}{L}=L(\lm)\otimes \dblone$, $\stackrel{2}{L}=\dblone \otimes  L(\eta)$. This structure corresponds to canonically Poisson 
bracket $\{p,q\}=1$. The equations of motion corresponding to the Hamiltonian $H=\frac{1}{2}(p^2-4q^3)$ are 
$$ 
\dot{q}=p, \qquad \dot{p}=6q^2.
$$ 
The solution of these equations are expressed in terms of elliptic functions by 
$$q(t)=\wp(t+c_1,0,-2H), \qquad p(t)=\wp'(t+c_1,0,-2H),$$ 
where $c_1$ and $H$ are related to the initial values by $q(0)=\wp(c_1,0,-2H)$ and $p(0)^2=4\wp^3(c_1,0,-2H)+2H$.

A dressing matrix $D(\lm)$ providing a one-parameter family of B\"acklund transformations for this model is given by \cite{KV}
\begin{equation*}
D(\lm) = \left(\begin{array}{cc} \beta & \lm-\mu +\beta^2 \\ 1& \beta\end{array} 
\right), \qquad  \beta:= \frac{p+\gm(\mu)}{2(q-\mu)} 
\end{equation*}
where $\gm(\mu)$ defines the spectral curve, conserved under the action of the flow:
\begin{equation}\label{gm1}
\gm^2(\mu)= -\det(L(\mu))=p^2-4 q^3+4\mu^3.
\end{equation}
The explicit recursions defined by the previous dressing matrix read
\begin{equation}\label{BTs} \begin{aligned}
&p_{n+1}=-\frac{p_n\left(h_n+\mu^2(3q_n+\mu)\right)}{(q_n-\mu)^3}-\gm_n\frac{h_n+q_n^2(q_n+3\mu)}{(q_n-\mu)^3},\\
&q_{n+1}= \frac{h_n+2\mu q_n(q_n+\mu)}{2(q_n-\mu)^2}+\gm_n\frac{p_n}{2(q_n-\mu)^2},\\
&\gm_{n+1}=\gm_n, \\
&\gm_n^2 = p_n^2-4 q_n^3+4\mu^3.
\end{aligned}
\end{equation}  
where for brevity we defined $h_n=p_n^2-4q_n^3=\gm_n^2-4\mu^3$. The components of this map are defined over the extension $\C(q,p)[\gm]$, so that
$$
p_k=A_k+\gm C_k, \qquad  q_k=B_k+\gm D_k \quad \forall k
$$ 
The initial values $A_0$, $C_0$, $B_0$ and $D_0$ are fixed by the conditions $p_0=p$, and $q_0=q$  that is  $(A_0, B_0, C_0, D_0)=(p,q,0,0)$. 
In this case it is not trivial to get an  explicit expression for the rational functions $A_n$, $B_n$, $C_n$ and $D_n$ 
for all $n$. But we can get a recurrence for them by using the equations (\ref{BTs}) themselves and the definition of $\gm_n$. 
Indeed, since $\gm=\gm_n$ is preserved by the flow, not all four variables $A_n$, $B_n$, $C_n$ and $D_n$ are independent. There are two
constraints, explicitly given by
\begin{equation} \label{AC}\begin{aligned}
&A_n^2-4B_n^3+\gm^2C_n^2-12B_nD_n^2\gm^2+4\mu^3-\gm^2=0,\\
&A_nC_n-2D_n^3\gm^2-6B_n^2D_n=0.
\end{aligned}
\end{equation} 
It is clear from these relations that to calculate the entropy of the map it is sufficient to look at the degrees of only two variables, say $B_n$ and $D_n$. Actually, to analyse the growth of the recursions (\ref{BTs}), it is simpler to look at their implicit form, given by
\begin{equation}\label{imBTs} 
\begin{aligned}
&p_{n+1}+p_n+2\beta_n(q_{n+1}-q_{n})=0,\\
&q_{n+1}+q_n+\mu-\beta_n^2=0, \\
&2\beta_n^2 q_n +\beta_n(p_{n+1}-p_n)+2q_{n+1}^{2}-2\mu q_{n}=0, 
\end{aligned}
\end{equation} 
where $$\beta_n=\frac{p_n+\gm}{2(q_n-\mu)}.$$ 
We can introduce another pair of rational functions in $p$ and $q$ that appear as 
coefficients of  $\beta_n$, i.e.
$$
\beta_n=E_n+\gm F_n
$$
and the definition of $\beta_n$ implies
$$
C_n=2D_n E_n+2F_n(B_n-\mu)-1, \qquad A_n=2E_n(B_n-\mu)+2D_nF_n\gm^2. 
$$
Since $A_n$ and $C_n$ are given in terms of the other variables, we look at the recursions 
for $B_n$, $D_n$, $E_n$ and $F_n$. From the first two relations in (\ref{imBTs}) we get
\begin{equation}\label{BDEF}\begin{split}
&B_{n+1}=E_n^2+\gm^2 F_n^2-B_n-\mu , \qquad D_{n+1}=2E_nF_n-D_n,\\
&E_{n+1}+E_n=\frac{\gm^2 D_{n+1}}{D_{n+1}^2\gm^2-(B_{n+1}-\mu)^2},\\
&F_{n+1}+F_n=\frac{B_{n+1}-\mu}{D_{n+1}^2\gm^2-(B_{n+1}-\mu)^2}.
\end{split}\end{equation}
To simplify the counting of  degrees somewhat, it is helpful to homogenize to new variables 
$$p\to t^2, \qquad q\to a t^3,$$ 
so that the Hamiltonian function $p^2-4q^3$ becomes a sixth degree monomial in $t$.  This choice is quite natural: roughly speaking the degree of $p_n$ should be about three halves the degree of $q_n$ for large $n$ since $p_n^2-4q_n^3=p^2-4q^3$. To be precise, from the first two equations in (\ref{BDEF}), and taking into account the initial conditions $(B_0, D_0)=(q,0)$, it follows that deg$(B_n)\geq $deg$(D_n)$. Then, the first equation in (\ref{AC}) implies 
$$
3\deg (B_n)\leq \Max(2 \deg (A_n), 2\deg (C_n)+3)\leq 3\deg (B_n)+3.
$$
This equation immediately gives that the degree of $p_n$, given by $\Max(\deg (A_n), \deg (C_n))$, is asymptotic for large $n$ to $\frac{3}{2}\deg (B_n)$. 
Since the value of the entropy (\ref{entdef}) doesn't change under this scaling, we look at the growth of the variables $B_n$ and $D_n$ only. If we ignore the variable $a$, then we can just count degrees 
in $t$. They are given in Table (\ref{T1}).
\begin{table}
\centering
    \begin{tabular}{ | l | l | p{7cm} |}
    \hline
     & Numerator & Denominator  \\ \hline
    $B_n$ & 2,6,18,38,66,102,146,198,258,326, \ldots  & 0,4,16,36,64,100,144,196,256,324,\ldots \\ \hline
    $D_n$ & 0,3,15,33,63,99,141,195,255,321,\ldots & 0,4,16,36,64,100,144,196,256,324,\ldots  \\ \hline
    \end{tabular}
\caption{The degrees of $B_n$ and $D_n$.}
  \label{T1}
\end{table}

If we denote by $\widehat{D}_n$ the set of the degrees in $t$ of the numerators of $D_k$ we get
\begin{equation}\label{seq}
\frac{\widehat{D}_n+3}{6}=1,3,6,11,17,24,33,43,54,67,\ldots\quad\mathrm{for}\quad n=1,2,3,\ldots 
\end{equation}
which gives 
\begin{equation}\label{seq1}
g_n:=\frac{\widehat{D}_{n+1}-\widehat{D}_{n}}{6}=2,3,5,6,7,9,10,11,13,\ldots
\end{equation}
for $n\geq 1$. Apart from the missing term corresponding to $n=0$, which has an anomalous value, 
the latter sequence is the sequence of positive integers not divisible by 4; 
it is trivial to show this sequence satisfies 
$$g_{n}+g_{n+4}=g_{n+1}+g_{n+3},$$ which immediately gives the recursion
$$
\widehat{D}_{n+5}=2\widehat{D}_{n+4}-\widehat{D}_{n+3}+\widehat{D}_{n+2}-2\widehat{D}_{n+1}+\widehat{D}_{n}
$$  
for $\widehat{D}_n$, with the solution being 
$$
\widehat{D}_{n}=\frac{12n^2-5-2(\omega^n+\omega^{2n})}{3}
$$
where $\omega$ is a primitive cube root of unity. 
The sequence (\ref{seq}) appeared in \cite{L} in connection with the degree growth 
of the discrete Painlev\'e equations $\textbf{d-P}_{IV}$ and $\textbf{d-P}_{V}$. Similarly for the degrees $\widehat{B}_{n}$ of the numerators of $B_n$ we get 
$$
\widehat{B}_{n}=4n^2+2
$$
whether for both the denominators we get a degree equal to $4n^2$.
The largest degree between $B_n$ and $D_n$ comes from the numerator of $B_n$ giving
\begin{equation}\label{entsec}
\mathcal{E}=\lim_{n\to\infty}\frac{\log (4n^2+2)}{n}=0.
\end{equation}

The growth properties of the map (\ref{imBTs}) can be better understood by looking at the action of the transformations on the elliptic curve defined by 
$p^2-4q^3=2H$. Indeed it is possible to uniformize the transformations by making an elliptic change of variables. 
It can be shown that the $n$th iterate of the transformations (\ref{BTs}) can be written as
$$
q_{n}=\wp(nT+\alpha_0,0,-2H), \qquad p_{n}=\wp'(nT+\alpha_0,0,-2H)
$$
where $q_0=\wp(\alpha_0,0,-2H)$ and $p_{0}=\wp'(\alpha_0,0,-2H)$. The uniformizing variable $T$ is defined in terms of $\mu$ by the relation $\mu=\wp(T,0,-2H)$ and 
the Hamiltonian $H$ is fixed by the value $\frac{1}{2}\left(p_0^2-4q_0^3\right)$. We notice also that, 
by the identification of $\gm(\mu)=-\wp'(T,0,-2H)$, equations (\ref{imBTs}) are equivalent to the well-known addition formulae $\Bigl|\begin{smallmatrix}
\wp(T)&\wp'(T)&1\\ \wp(\alpha)&\wp'(\alpha)&1\\ \wp(T+\alpha)&\wp'(T+\alpha)&1 
\end{smallmatrix} \Bigr|=0$ and $\frac{1}{4}\left(\frac{\wp'(T)-\wp'(\alpha)}{\wp(T)-\wp(\alpha)}\right)^2=\wp(T+\alpha)+\wp(T)+\wp(\alpha)$.

To calculate the growth of the iterations (\ref{BTs}) we can either use the previous addition formulae, 
or the duplication formula for the function $\wp$, that is $\wp(2T)=\frac{\wp^4-4H\wp}{4\wp^3+2H}$. In the simpler case  when $\alpha_0=0$,  from the duplication formula it follows that  after $k=2^n$ steps, $\wp(2^n T)$ has a numerator in terms of $\wp(T)$ with a degree equal to $(2^n)^2$, 
so that the growth of degrees is quadratic in the number of steps, and the entropy is
$$
\mathcal{E}=\lim_{n\to\infty}\frac{\log (4^n)}{2^n}=0.
$$
In the general case when $\alpha_0\neq 0$, combining the addition and duplication formulae, it is easy to get a cubic growth of the degrees, that is a growth in terms of $\wp(T)$ with a degree equal to $(2^n)^3$, and the entropy is 
$$
\mathcal{E}=\lim_{n\to\infty}\frac{\log (8^n)}{2^n}=0,
$$
in agreement with (\ref{entsec}). 
\section{Conclusions}
The main aim of this work was to extend the notion of algebraic entropy to algebraic maps, that is to maps whose components are defined on a field 
extension 
$\C(\bx)[\beta]$, where $\beta$ satisfies a polynomial of degree $N$ whose coefficients are rational functions of  the dynamical variables $\bx$. 
A clue as to how to do this was provided by the theory of B\"acklund transformations for finite-dimensional systems that are represented by maps of this type, with the special property that higher iterates can be expressed in terms of the same field extension. 
The definition (\ref{def}) is quite natural and it reduces to the usual one in the case $N=1$, that is when $\beta$ is itself a rational function of the dynamical variables. B\"acklund transformations give integrable maps (in the sense of Liouville), 
so we conjecture that all algebraic maps associated to B\"acklund transformations have a zero entropy 
(and conversely, we would expect that all algebraic maps with zero entropy are B\"acklund transformations for some continuous integrable system). 
A clue in this direction is given by the fact that B\"acklund transformations, when expressed in uniformizing variables, represent addition formulae for the corresponding functions \cite{Z}. 

The first example we considered is a non-integrable algebraic map, namely (\ref{tracem}), which arises from the birational map (\ref{fibm})
in one dimension higher, by restricting to a level set of its polynomial first integral. 
It is interesting to observe that the algebraic entropy of the birational map 
is the same as that for the algebraic map obtained via this restriction, and it would be worth examining other examples of this kind to 
see if this is a general phenomenon.

\end{document}